\RequirePackage{fix-cm}
\documentclass[smallextended]{svjour3}       
\smartqed  
\usepackage[T1]{fontenc}
\usepackage[latin9]{inputenc}
\usepackage{graphicx}
\usepackage{appendix}
\usepackage{tikz}
\usepackage{amsmath}
\usepackage{amssymb}
\usepackage{bm}
\usepackage[unicode=true,pdfusetitle,
 bookmarks=true,bookmarksnumbered=false,bookmarksopen=false,
 breaklinks=false,pdfborder={0 0 0},pdfborderstyle={},backref=false,colorlinks=true]
 {hyperref}
\usepackage{tabularx}
\usepackage{makecell}

\allowdisplaybreaks[4]

\begin{document}

\title{Even- and odd-orthogonality properties of the Wigner
D-matrix and their metrological applications}

\author{Wei Zhong \and Lan Zhou \and Cui-Fang Zhang \and Yu-Bo Sheng}


\institute{
W. Zhong \at
Institute of Quantum Information and Technology, Nanjing University of Posts and Telecommunications, Nanjing 210003, China\\
National Laboratory of Solid State Microstructures, Nanjing University, Nanjing 210093, China\\
\email{zhongwei1118@gmail.com}
\and
L. Zhou \at
School of Science, Nanjing University of Posts and Telecommunications, Nanjing 210003, China\\
\and
C. F. Zhang  \at
Institute of Quantum Information and Technology, Nanjing University of Posts and Telecommunications, Nanjing 210003, China\\
\and
Y. B. Sheng \at
College of Electronic and Optical Engineering, Nanjing University of Posts and Telecommunications, Nanjing 210003, China\\
\email{shengyb@njupt.edu.cn}
}

\date{Received: date / Accepted: date}

\maketitle

\begin{abstract}
The Wigner D-matrix is essential in the course of
angular momentum techniques. We here derive the new even- and odd-orthogonality
properties of the Wigner D-matrix which was yet
to be demonstrated in textbooks and also apply them to identifying
optimal measurements for linear phase estimation based on two-mode
optical interferometry with two specific quantum states.
\end{abstract}

\maketitle

\section{Introduction}

The quantum theory of angular momentum, as developed principally by
Eugene Wigner \cite{Wigner1967Book}, has become an indispensable
discipline for quantitative study of problems in atomic physics, molecular
physics, nuclear physics and solid state physics \cite{Biedenhar1984Book}.
Angular momentum theory is of greater importance in the theory of
atomic and molecular physics and other quantum problems involving
rotational symmetry. It is naturally applied to represent the physical
state of a quantum system, due to the inherent connection between
the properties of elementary particles to the structure of Lie group
and Lie algebra. Angular momentum operators are defined as the generators
of SU(2) and SO(3) groups, which consists of a mathematical
structure of a Lie algebra.

To quantitatively represent a state of quantum system
of $N$ qubits, the routine means is to use the collective angular
momentum operators $J_{\xi}=\sum_{_{i=1}}^{N}\sigma_{i\xi}/2$ $\left(\xi=x,y,z\right)$
with $\sigma_{i\xi}$ the Pauli matrix of the $i$th qubit, satisfying
the following commutation relations $\left[J_{i},J_{j}\right]=i\epsilon_{ijk}J_{k}$$\ \left(i,j,k=\left\{ x,y,z\right\} \right)$
with $\epsilon_{ijk}$ the Levi-Civita symbol. The basis space is
spanned by the common eigenstates $\vert j,m\rangle$ of the operators
$J^{2}=J_{x}^{2}+J_{y}^{2}+J_{z}^{2}$ and $J_{z}$ with the eigenvalues
$j\left(j+1\right)$ and $m=-j,-j+1,\ldots,j$, respectively, where
the quantum number $j$ in the spherical basis $\vert j,m\rangle$
is given by $j=0,1/2,1,3/2,2,\ldots$ for SU(2), and $j=0,1,2,\ldots$
for SO(3). A general $3$-dimensional rotation operator can be conveniently
described in terms of the three Euler angles $\alpha$, $\beta$ and
$\gamma$ $\left(0<\alpha<2\pi,0<\beta<\pi,0<\gamma<2\pi\right)$
as
\begin{eqnarray}
R\left(\alpha,\beta,\gamma\right) & \equiv & e^{-i\alpha J_{z}}e^{-i\beta J_{y}}e^{-i\gamma J_{z}},
\end{eqnarray}
In the basis of $\left\{ \vert j,m\rangle\right\} $, the rotation
operator $R$ can be represented as a square matrix of dimension $2j+1$
with general element
\begin{eqnarray}
D_{m^{\prime}\!,m}^{j}\!\left(\alpha,\beta,\gamma\right) & \!=\! & \langle j,m^{\prime}\vert R\left(\alpha,\beta,\gamma\right)\vert j,m\rangle\!=\!e^{-im^{\prime}\!\alpha}d_{m^{\prime}\!,m}^{j}\!\left(\beta\right)e^{-im\gamma},\quad\label{eq:D-matrix}
\end{eqnarray}
which is the so-called Wigner D-matrix introduced in 1927 by Eugene
P. Wigner. The Wigner D-matrix is a matrix in an irreducible representation
of the groups SU(2) and SO(3). In above, the matrix with general element
\begin{eqnarray}
d_{m^{\prime},m}^{j}\left(\beta\right) & \equiv & \langle j,m^{\prime}\vert e^{-i\beta J_{y}}\vert j,m\rangle,
\end{eqnarray}
 is known as the Wigner (small) $d$-matrix \cite{Wigner1959Book}.
Apparently, the Wigner D-matrix reduces to the Wigner (small) $d$-matrix
when we let $\alpha=\gamma=0$.

Wigner rotation matrix has been proposed for over
a century and its relevant properties are almost comprehensively written
into textbooks \cite{Wigner1959Book,Biedenhar1984Book}, for instance,
the orthogonality relation of the Wigner D-matrix
\begin{eqnarray}
\sum_{\mu=-j}^{j}D_{m^{\prime},\mu}^{j}\left(\alpha,\beta,\gamma\right)D_{m,\mu}^{j}\left(\alpha,\beta,\gamma\right)^{\ast} & = & \delta_{m^{\prime},m}.\label{eq:D-orthogonality}
\end{eqnarray}
In this paper, we introduce two new orthogonality relations of the
Wigner D-matrix which was yet to be demonstrated
in textbooks. We first present a brief derivation of them by simply
specifying a quantum state $\vert\psi\rangle$ in an ansatz space
associating with zero-value expectation of $J_{z}$ for $\vert\psi\rangle$,
i.e., $\langle J_{z}\rangle_{\psi}=0$. We then apply these orthogonality
relations to two-mode optical interferometry to identify optimal measurements
to access the ultimate sensitivity of phase estimation with two specific
states.

This paper is organized as follows. In Sect. 2, we first explicitly
derive the two new orthogonality relations of the Wigner
D-matrix. In. Sect. 3, we invoke these relations to identify the
optimal measurements for Mach-Zehnder interferometry with NOON and
entangled coherent states. Finally, the conclusions are given in Sect.
4.

\section{Even- and odd-orthogonality properties of the Wigner D-matrix}

We first present the two novel even- and odd-orthogonality
properties in terms of the Wigner D-matrix that we would like to derive
in the following. For odd $N$, we have the following identities
\begin{eqnarray}
\sum_{k=\left\{ {\rm even,odd}\right\} }D_{j-k,m}^{j}\left(\alpha,\frac{\pi}{2},\gamma\right)D_{j-k,m^{\prime}}^{j}\left(\alpha,\frac{\pi}{2},\gamma\right)^{\ast} & = & \frac{1}{2}\delta_{m,m^{\prime}},\label{eq:even-odd-orthogonality}
\end{eqnarray}
where $k$ runs over the even or odd numbers from $0$ to $N$. As
such, we call the above relations as even- and odd-orthogonality properties
of the Wigner D-matrix. The above identities also hold for even $N$
when $m\neq0$, but when $m=0$ they become
\begin{eqnarray}
\sum_{k={\rm \left\{ even,odd\right\} }}D_{j-k,0}^{j}\left(\alpha,\frac{\pi}{2},\gamma\right)D_{j-k,m^{\prime}}^{j}\left(\alpha,\frac{\pi}{2},\gamma\right)^{\ast} & = & \delta_{0,m^{\prime}},\label{eq:even-odd-0-orthogonality}
\end{eqnarray}
up to absence of a fact of $1/2$. Obviously, Eqs.~\eqref{eq:even-odd-orthogonality}
and \eqref{eq:even-odd-0-orthogonality} are different
from Eq.~\eqref{eq:D-orthogonality},
since one can obtain Eq.~\eqref{eq:D-orthogonality}
with $\beta=\pi/2$ by summarizing Eqs.~\eqref{eq:even-odd-orthogonality}
and \eqref{eq:even-odd-0-orthogonality}, while
the opposite is not possible. The above expressions can be simplified
to a form where the conjugation operation can be removed for $\alpha=\gamma=0$
since the resulting $d$-matrix elements as $D_{m^{\prime}\!,m}^{j}\!\left(0,\beta,0\right)=d_{m^{\prime}\!,m}^{j}\!\left(\beta\right)$
are real.

To derive the above identities, we first recall the
orthogonality property of the Wigner D-matrix given by Eq.~\eqref{eq:D-orthogonality}
which can be directly derived from the fact that the normalization
condition of a quantum state $\vert\psi\rangle$ is still sustained
after a rotation operation $R\left(\alpha,\beta,\gamma\right)$, suggesting
that $\langle\psi\vert R^{\dagger}\left(\alpha,\beta,\gamma\right)R\left(\alpha,\beta,\gamma\right)\vert\psi\rangle=\langle\psi\vert\psi\rangle=1$
due to $R^{\dagger}R=\bm{1}$. Assume the total number of qubits
of the system is fixed as $N$, thus identifying $j=N/2$. Thus a
generic pure state can be expanded in terms of the basis of $\left\{ \vert j,m\rangle\right\} $
as $\vert\psi\rangle=\sum_{m}C_{m}\vert j,m\rangle$, associating
with $\sum_{m}\left|C_{m}\right|^{2}=1$. We denote the state after
the rotation operation $R\left(\alpha,\beta,\gamma\right)$ by $\vert\tilde{\psi}\rangle\equiv R\left(\alpha,\beta,\gamma\right)\vert\psi\rangle$.
It can also be expanded as $\vert\tilde{\psi}\rangle=\sum_{\mu}\tilde{C}_{\mu}\vert j,\mu\rangle$
where the expanded coefficients is given by $\tilde{C}_{\mu}\equiv\sum_{m}D_{\mu,m}^{j}\left(\alpha,\beta,\gamma\right)C_{m}$,
satisfying the normalization condition $\sum_{\mu}\vert\tilde{C}_{\mu}\vert^{2}=1$.
With the normalization $\sum_{\mu}\vert\tilde{C}_{\mu}\vert^{2}=\sum_{m}\vert C_{m}\vert^{2}=1$,
the orthogonality relation of Eq.~\eqref{eq:D-orthogonality} can
be obtained.

The procedure used in above can be straightforwardly
extended to derive Eqs.~\eqref{eq:even-odd-orthogonality} and \eqref{eq:even-odd-0-orthogonality}.
Here we make two restrictions. First, the pure state $\vert\psi\rangle$
is assumed to be in an ansatz subspace satisfying $C_{m}=C_{-m}$,
implying that $\langle J_{z}\rangle_{\psi}=0$, which encompasses
an amount of quantum states implemented in experiments \cite{Caves1981PRD,Campos2003PRA,Pezze2008PRL,Huver2008PRA,Hofmann2009PRA,Lucke2011Science,Anisimov2010PRL,Liu2013PRA,Pezze2013PRL,Froewis2014NJP,Zhong2017PRA,Zhong2020SC,Zhong2021PRA,Lang2013PRL}.
Note that this restriction we assume here is simply convenient for
our derivation, but the final result does not depend on what type
of quantum state is used. Second, the rotation angle $\beta$ is specified
by $\beta=\pi/2$, giving $R\left(\alpha,\pi/2,\gamma\right)$. This
is a very reasonable assumption. If let $\alpha=\gamma=0$, thus the
rotation operation is $R_{y}\left(\pi/2\right)\equiv R\left(0,\pi/2,0\right)$,
which can be implemented to characterize a class of quantum coherent
operations, such as a balanced beam splitter in linear optical system
or a $\pi/2$-pulse in atomic system. Under these constraints, the
expansion coefficients for the state after the rotation is given by
\begin{eqnarray}
\tilde{C}_{\mu} & = & \sum_{m=-j}^{j}\!C_{m}\,D_{\mu,m}^{j}\!\left(\alpha,\frac{\pi}{2},\gamma\right).
\end{eqnarray}
With the help of $C_{m}=C_{-m}$, it can be expressed as the following
form
\begin{eqnarray}
\tilde{C}_{\mu} & = & \begin{cases}
\sum\limits _{m=1/2}^{j}C_{m}\left[D_{\mu,-m}^{j}\left(\alpha,\frac{\pi}{2},\gamma\right)+D_{\mu,m}^{j}\left(\alpha,\frac{\pi}{2},\gamma\right)\right], & \text{for odd }N,\\
\sum\limits _{m=1}^{j}C_{m}\left[D_{\mu,-m}^{j}\left(\alpha,\frac{\pi}{2},\gamma\right)+D_{\mu,m}^{j}\left(\alpha,\frac{\pi}{2},\gamma\right)\right]+C_{0}D_{\mu,0}^{j}\left(\alpha,\frac{\pi}{2},\gamma\right), & \text{for even }N.
\end{cases}
\end{eqnarray}
By employing the identity $D_{\mu,-m}^{j}\left(\alpha,\frac{\pi}{2},\gamma\right)=\left(-1\right)^{-j-\mu}D_{\mu,m}^{j}\left(\alpha,\frac{\pi}{2},\gamma\right)$
\cite{Wigner1967Book,Biedenhar1984Book}, the above expression can
be further represented as
\begin{eqnarray}
\tilde{C}_{\mu} & = & \begin{cases}
\sum\limits _{m=1/2}^{j}C_{m}\left[1+\left(-1\right)^{-j-\mu}\right]D_{\mu,m}^{j}\left(\alpha,\frac{\pi}{2},\gamma\right), & \text{for odd }N,\\
\sum\limits _{m=1}^{j}C_{m}\left[1+\left(-1\right)^{-j-\mu}\right]D_{\mu,m}^{j}\left(\alpha,\frac{\pi}{2},\gamma\right)+C_{0}D_{\mu,0}^{j}\left(\alpha,\frac{\pi}{2},\gamma\right), & \text{for even }N.
\end{cases}\label{eq:enpandcoefficients}
\end{eqnarray}

For simplicity, we below take the case of odd $N$
for example. With Eq.~\eqref{eq:enpandcoefficients} and by setting
$k\equiv j+\mu$, we have
\begin{eqnarray}
\sum_{\mu=-j}^{j}\left|\tilde{C}_{\mu}\right|^{2} & = & \sum_{m,m^{\prime}=1/2}^{j}2C_{m}C_{m^{\prime}}^{\ast}\left\{ \sum_{k=0}^{N}\left[1+\left(-1\right)^{-k}\right]D_{j-k,m}^{j}\left(\alpha,\frac{\pi}{2},\gamma\right)D_{j-k,m^{\prime}}^{j}\left(\alpha,\frac{\pi}{2},\gamma\right)^{\ast}\right\} \nonumber \\
 & = & \sum_{m,m^{\prime}=1/2}^{j}2C_{m}C_{m^{\prime}}^{\ast}\times\left[2\sum_{k={\rm even}}D_{j-k,m}^{j}\left(\alpha,\frac{\pi}{2},\gamma\right)D_{j-k,m^{\prime}}^{j}\left(\alpha,\frac{\pi}{2},\gamma\right)^{\ast}\right].
\end{eqnarray}
Obviously it suggests the relation of Eq.~\eqref{eq:even-odd-orthogonality}
with $k=$ even numbers from $0$ to $N$ must be satisfied due to
the normalization $\sum_{\mu=-j}^{j}\vert\tilde{C}_{\mu}\vert^{2}=\sum_{m=-j}^{j}\left|C_{m}\right|^{2}=\sum_{m=1/2}^{j}\!2\left|C_{m}\right|^{2}=1$.
Thus the relation of Eq.~\eqref{eq:even-odd-orthogonality} with
$k=$ odd numbers from $0$ to $N$ can be obtained by taking the
difference between Eq.~\eqref{eq:D-orthogonality} with $\beta=\pi/2$
and Eq.~\eqref{eq:even-odd-orthogonality}. Similarly, the derivation
for the case of even $N$ just follows the same procedure used in
above derivation, but minding the term of $m=0$, which needs to be
treated solely in the square sum calculation.

\section{Optimal measurements for linear phase estimation}
\begin{center}
\begin{figure}[t]
\centering{}\includegraphics[scale=0.13]{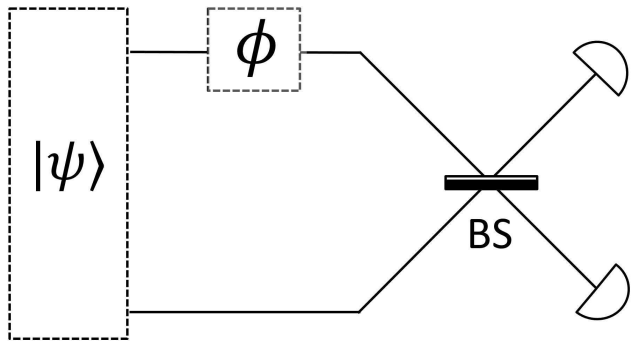}\caption{(Color online) Schematic of linear phase estimation based on two-mode
optical interferometry. \label{fig:set-up}}
\end{figure}
\par\end{center}

To view the effectiveness of even- and odd-orthogonality properties
obtained above, we apply them into the linear phase estimation problem
to identify optimal measurement for saturating the ultimate sensitivity
limit in Mach-Zehnder interformetry with different probe states. A
generic MZI is composed of two balanced beam splitters $B_{i}$ $(i=1,2)$
and a phase shifting $U_{\phi}$ with $\phi$ to be estimated. Here
we identify the state after the first beam splitter $B_{1}$ as the
probe state $\rho\equiv B_{1}\rho_{{\rm in}}B_{1}^{\dagger}$ with
$\rho_{{\rm in}}$ the input state injecting at the input port of
the interferometer (see Fig. \ref{fig:set-up}). Under
$U_{\phi}$, the probe state evolves into a parametric state as $\rho\left(\phi\right)=U_{\phi}\rho U_{\phi}$.
Finally, a measurement is performed at the output port of the interferometer
where the output state reads $\rho_{{\rm out}}\left(\phi\right)=B_{2}\rho\left(\phi\right)B_{2}^{\dagger}$,
and then the true value of the phase shift is inferred from the measurement
outcomes.

Consider a general measurement described by a positive-valued
measure $\bm{M}=\left\{ M_{\chi}\right\} $, with $\chi$ the results
of measurement. Based on $\bm{M}$, the sensitivity of an unbiased
phase estimator $\phi_{{\rm est}}$ is limited by the classical Cram\'er-Rao
bound $\Delta^2\phi_{{\rm est}}\geq\left[\upsilon F\left(\rho_{{\rm out}}\left(\phi\right)\vert\bm{M}\right)\right]^{-1}$,
where $\upsilon$ is the repetitions of the experiments and $F$ is
the classical Fisher information (CFI) defined by \cite{Helstrom1976Book,Holevo1982Book}
\begin{equation}
F=\sum_{\chi}\frac{1}{p\left(\chi\vert\rho_{{\rm out}}\left(\phi\right)\right)}\left[\frac{\partial p\left(\chi\vert\rho_{{\rm out}}\left(\phi\right)\right)}{\partial\phi}\right]^{2},
\end{equation}
with $p\left(\chi\vert\rho_{{\rm out}}\left(\phi\right)\right)={\rm Tr}\left(M_{\chi}\rho_{{\rm out}}\left(\phi\right)\right)$
the probability of the outcome $\chi$ conditioned on the specific
value of $\phi$ contained in $\rho_{{\rm out}}\left(\phi\right)$.
It is well known that such an accessible sensitivity bound is achieved
by the maximum likelihood estimator for sufficiently large $\upsilon$
with Bayesian estimation methods \cite{Krischek2011PRL,Olivares2009JPB,Uys2007PRA}.
A quantum analog of Cram\'er-Rao bound is identified by optimizing
the measurements over the positive-valued measure operators $\bm{M}$,
such that $\Delta\phi_{{\rm est}}\geq\left(\upsilon H\right)^{-1}$
with $H=\sup_{\bm{M}}\left(F\right)$ denoting the quantum Fisher
information (QFI) \cite{Braunstein1994PRL,Lu2012PRA}. For pure states
$\rho\left(\phi\right)=\vert\psi\left(\phi\right)\rangle\langle\psi\left(\phi\right)\vert$
with $\vert\psi\left(\phi\right)\rangle=U_{\phi}\vert\psi\rangle$,
the QFI can be explicitly expressed as
\begin{equation}
H=4\left(\langle\partial_{\phi}\psi\left(\phi\right)\vert\partial_{\phi}\psi\left(\phi\right)\rangle-\left|\langle\psi\left(\phi\right)\vert\partial_{\phi}\psi\left(\phi\right)\rangle\right|^2\right),\label{eq:QFI}
\end{equation}
with $\partial_{\phi}\psi\left(\phi\right)\equiv\partial\psi\left(\phi\right)/\partial\phi$
being the derivative of $\psi\left(\phi\right)$ with respect to $\phi$.
If we assume that the phase parameter $\phi$ here is imprinted via
a unitary operation $U_{\phi}=\exp\left(-iG\phi\right)$ with $G$
the generator, then the expression of Eq.~\eqref{eq:QFI} can be
simplified as four times the variance of the generator $G$ for the
pure state $\vert\psi\rangle$, equivalently
\begin{equation}
H=4\left\langle \Delta^{2}G\right\rangle _{\psi}=4\big(\left\langle G^{2}\right\rangle _{\psi}-\left\langle G\right\rangle _{\psi}^{2}\big).\label{eq:QFI_var}
\end{equation}
Note that, in the setting as mentioned above, the QFIs for the states
before and after the second beams splitter are identical due to the
invariance property of the QFI under a phase-independent unitary evolution
\cite{Helstrom1976Book,Holevo1982Book}. We below take the CFI and
QFI as the two figure of merits, as they are widely
used to evaluate the performance of various metrological applications
\cite{Ozaydin2015SR,Ozaydin2020OQE,Wang2020Met,Zhong2013PRA,Cai2020QIP,Guo2022PRA,Zhong2020SC,Zhong2017PRA,Zhong2014CPB,Taylor2016PR,Pezze2018RMP,Polino2020review,Sparaciari2015JOSAB,Sparaciari2016PRA}.
Hence, an efficient way to identify a measurement being optimal for
a quantum phase estimation task is to test whether the QFI and the
CFI with respect to a specific measurement are identical \cite{Braunstein1994PRL,Zhong2014JPA}.

In what follows, the probe states of interest here are restricted
to NOON \cite{Rarity1990PRL,Lee2002JOP} and entangled coherent (EC)
states \cite{Joo2011PRL} (see below for definitions). The measurement
of interest here is a double-photon-counting (DPC) detection, which
can be represented as the projection operator $M=\left\{ \vert n_{a},n_{b}\rangle\langle n_{a},n_{b}\vert\right\} $
\cite{Pezze2008PRL,Pezze2013PRL,Lang2013PRL,Zhong2017PRA} with the
pairs of outcomes ($n_{a},n_{b}$) recording the photon numbers detected
at the two output ports of the interferometer. Thus, to implement
this measurement, a high-efficiency photon-number-resolving detector
is demanded in experiments \cite{Divochiy2008NP,Sahin2013APL}.

\subsection{NOON states}

Let us first consider the NOON state as the probe state, which can
be expressed as $\left|N::0\right\rangle =\left(\vert N0\rangle+\vert0N\rangle\right)/\sqrt{2}$
with $\vert N0\rangle$ $\left(\vert0N\rangle\right)$ denoting the
$N$ photons (null) in mode $a$ and null ($N$ photons) in mode $b$,
respectively. The operation of phase shifting is assumed to be expressed
as $U\left(\phi\right)=\exp\left(-ia^{\dagger}a\phi\right)$, where
$a^{\dagger}\left(a\right)$ stands for the creation (annihilation)
operator of the corresponding mode. Hence, the NOON state under $U\left(\phi\right)$
evolves into a parametric one $\left|N::0,\phi\right\rangle \equiv U\left(\phi\right)\vert N::0\rangle$.
Such a strategy is expected to be possibly achieving a Heisenberg-scaling
sensitivity $\Delta\phi\sim N^{-1}$ \cite{Rarity1990PRL,Lee2002JOP},
that is $H_{{\rm NOON}}=N^{2}$ \cite{Zhong2021PRA}.

To access this limit, we consider the DPC measurement. The probability
of detecting the photon numbers $\left(n_{a},n_{b}\right)$ at the
two-output ports is given by
\begin{equation}
p\left(n_{a},n_{b}\vert N::0,\phi\right)=\left|\langle n_{a},n_{b}\vert R_{y}\left(\frac{\pi}{2}\right)\left|N::0,\phi\right\rangle \right|^{2},
\end{equation}
where we have assumed the operation of the second BS to be $B_{2}=R_{y}\left(\pi/2\right)$
following the form commonly adopted in previous works \cite{Zhong2017PRA,Zhong2020SC,Zhong2021PRA,Guo2022PRA}.
For simplicity, we use the angular momentum technique (which is also
called Schwinger representation \cite{Yurke1986PRA}) to represent
the two-mode optical interferometry by defining $J_{x}=\left(a^{\dagger}b+ab^{\dagger}\right)/2$,
$J_{y}=\left(a^{\dagger}b-ab^{\dagger}\right)/2i$ and $J_{z}=\left(a^{\dagger}a-b^{\dagger}b\right)/2$.
Thus the NOON state can be rewritten by $\left|N::0\right\rangle =\left(\vert j,j\rangle+\vert j,-j\rangle\right)/\sqrt{2}$
with $j=N/2$, correspondingly, the parametric NOON state reads $\left|N::0,\phi\right\rangle =\left(e^{-ij\phi}\vert j,j\rangle+e^{ij\phi}\vert j,-j\rangle\right)/\sqrt{2}$.
Identifying $j^{\prime}=\left(n_{a}+n_{b}\right)/2\text{ and }m^{\prime}=\left(n_{a}-n_{b}\right)/2$,
one can obtain $p\left(n_{a},n_{b}\vert N::0,\phi\right)=p\left(j^{\prime},m^{\prime}\vert N::0,\phi\right)$.
Note that the irreducible representation of angular momentum requires
the two states spanned in different irreducible subspaces are orthogonal
in the sense that we have $j^{\prime}=j=N/2$. Hence, by setting $j-m^{\prime}\equiv k$,
the above expression can be further expressed as
\begin{eqnarray}
p\left(j,j-k\vert N::0,\phi\right) & = & \begin{cases}
\left[d_{j-k,j}^{j}\left(\frac{\pi}{2}\right)\right]^{2}2\cos^{2}\left(j\phi\right), & \text{even }k,\\
\left[d_{j-k,j}^{j}\left(\frac{\pi}{2}\right)\right]^{2}2\sin^{2}\left(j\phi\right), & \text{odd }k.
\end{cases}\label{eq:probability}
\end{eqnarray}
Submitting the above results into the definition of the CFI finally
yields
\begin{eqnarray}
F_{{\rm NOON}} & = & \sum_{n_{a},n_{b}}\frac{1}{p\left(n_{a},n_{b}\vert N::0,\phi\right)}\left[\frac{\partial p\left(n_{a},n_{b}\vert N::0,\phi\right)}{\partial\phi}\right]^{2}\nonumber \\
 & = & \sum_{k}\frac{1}{p\left(j,j-k\vert N::0,\phi\right)}\left[\frac{\partial p\left(j,j-k\vert N::0,\phi\right)}{\partial\phi}\right]^{2}\nonumber \\
 & = & \sum_{k={\rm even}}\left[d_{j-k,j}^{j}\left(\frac{\pi}{2}\right)\right]^{2}8j^{2}\sin^{2}\left(j\phi\right)+\sum_{k={\rm odd}}\left[d_{j-k,j}^{j}\left(\frac{\pi}{2}\right)\right]^{2}8j^{2}\cos^{2}\left(j\phi\right)\nonumber \\
 & = & N^{2},\label{eq:CFI_NOON}
\end{eqnarray}
where the last equality is obtained by employing the even- and odd-orthogonality
properties in terms of the Wigner $d$-matrix obtained
from Eq.~\eqref{eq:even-odd-orthogonality} by setting $\alpha=\gamma=0$.
Obviously, we find that $F_{{\rm NOON}}=H_{{\rm NOON}}$, in the sense
that the DPC detection is an optimal measurement for saturating the
Heisenberg limit expected by NOON states. Moreover, it is also shown
that the result of $F_{C,{\rm NOON}}$ is $\phi$-independent, suggesting
that the DPC detection is globally optimal over the full range of
phase shift values.

More importantly, from Eq.~\eqref{eq:probability}, it is shown that
the probability of detection only depends on a single variable $k$
(or equivalently $m^{\prime}$ due to the restriction of $N/2-m^{\prime}\equiv k$),
since the states over the whole evolution process are only confined
in the $j=N/2$ irreducible subspace. This means that a population
difference detection (i.e., measuring $J_{z}$) at the output ports
is sufficient for attaining the Heisenberg limit. Besides it also
has been demonstrated that parity detection exhibits an equivalent
performance as the DPC detection and population difference detection
(see also Appendix for a demonstration) \cite{Seshadreesan2013PRA,Zhong2014JPA}.

\subsection{Entangled coherent states}

Below, we consider the EC state as the probe state, which can be understood
as a superposition of NOON states with different photons,
\begin{eqnarray}
\vert{\rm EC}\rangle & = & \mathcal{N}_{\alpha}\left[\vert\alpha\rangle\vert0\rangle+\vert0\rangle\vert\alpha\rangle\right]=\sqrt{2}\mathcal{N}_{\alpha}\sum_{n=0}^{\infty}c_{n}\vert n::0\rangle,
\end{eqnarray}
with $\mathcal{N}_{\alpha}=1/\sqrt{2\left(1+e^{-\left|\alpha\right|^{2}}\right)}$
the normalization factor and $c_{n}=e^{-\vert\alpha\vert^{2}/2}\alpha^{n}/\sqrt{n!}$
the superposition coefficient of the coherent state. This state can
be generated by powering a coherent state into one input port mode
of a beam splitter and a coherent superposition of macroscopically
distinct coherent states into another input port \cite{Luis2001PRA}.
Note that a phase-averaging operation is required here in calculation
of the QFI due to the lack of an external reference beam in out setting
\cite{Jarzyna2012PRA}. Under the phase-averaging operation, the
EC states becomes a mixed state that consists of a statistical ensemble
of NOON states, that is,
\begin{eqnarray}
\rho_{{\rm EC}} & = & 2\mathcal{N}_{\alpha}^{2}\bigoplus_{n=0}^{\infty}\left|c_{n}\right|^{2}\left|n::0\right\rangle \left\langle n::0\right|.
\end{eqnarray}
We thus explicitly derive the QFI of the EC state as
\begin{eqnarray}
H_{{\rm EC}} & = & 2\mathcal{N}_{\alpha}^{2}\sum_{n=1}^{\infty}\left|c_{n}\right|^{2}H_{{\rm noon}}=2\mathcal{N}_{\alpha}^{2}\sum_{n=1}^{\infty}\left|c_{n}\right|^{2}n^{2},\label{eq:QFI_EC}
\end{eqnarray}
as a result of the summability of the QFI \cite{Helstrom1976Book,Fujiwara2001PRA}
and $H_{{\rm NOON}}=N^{2}$. For larger amplitude $\left|\alpha\right|\gg1$,
the expression of Eq.~\eqref{eq:QFI_EC} approximately reduced to
$H_{{\rm EC}}=\bar{N}+\bar{N}^{2}$ with the mean photon number $\bar{N}=\left|\alpha\right|^{2}$.
In contrast, it has been observed that a larger amount of $H_{{\rm EC}}^{J}=2\bar{N}+\bar{N}^{2}$
can be obtained when a common reference beam is involved \cite{Joo2011PRL}.
This indicates that higher estimation sensitivity could be attained
when the extra reference beam is established in current setting.

Now let us identify the CFI with the DPC detection for the EC probe
state. When powering EC states as the probe state, the probability
of detecting $n_{a}$ and $n_{b}$ at the output ports is given by
\begin{eqnarray}
p\left(n_{a},n_{b}\vert{\rm EC},\phi\right) & = & \left|\langle n_{a},n_{b}\vert R_{y}\left(\frac{\pi}{2}\right)\left|{\rm EC},\phi\right\rangle \right|^{2}\nonumber \\
 & = & 2\mathcal{N}_{\alpha}^{2}\sum_{n=0}^{\infty}\left|c_{n}\right|^{2}\left|\langle n_{a},n_{b}\vert R_{y}\left(\frac{\pi}{2}\right)\left|n::0,\phi\right\rangle \right|^{2}\nonumber \\
 & = & 2\mathcal{N}_{\alpha}^{2}\sum_{n=1}^{\infty}\left|c_{n}\right|^{2}p\left(n_{a},n_{b}\vert n::0,\phi\right)+4\mathcal{N}_{\alpha}^{2}\left|c_{0}\right|^{2},\label{eq:probability_nanb}
\end{eqnarray}
where $\left|{\rm EC},\phi\right\rangle $ corresponds to the parametric
EC state defined by $\left|{\rm EC},\phi\right\rangle =U\left(\phi\right)\vert{\rm EC}\rangle$
and in the last equality the second term is obtained by adopting Eq.~\eqref{eq:probability}
incorporating with $d_{0,0}^{0}\left(\pi/2\right)=1$. With Eq.~\eqref{eq:probability_nanb}
and transferring into the Schwinger representation, the CFI with the
DPC detection can be derived as follows
\begin{eqnarray}
F_{{\rm EC}} & = & \sum_{n_{a},n_{b}}\frac{1}{p\left(n_{a},n_{b}\vert{\rm EC},\phi\right)}\left[\frac{\partial p\left(n_{a},n_{b}\vert{\rm EC},\phi\right)}{\partial\phi}\right]^{2}\nonumber \\
 & = & \sum_{j\neq0}\sum_{\mu=-j}^{j}\frac{\left[2\mathcal{N}_{\alpha}^{2}\sum_{n=1}^{\infty}\left|c_{n}\right|^{2}\partial_{\phi}p\left(j,\mu\vert n::0,\phi\right)\delta_{j,n/2}\right]^{2}}{2\mathcal{N}_{\alpha}^{2}\sum_{n=1}^{\infty}\left|c_{n}\right|^{2}p\left(j,\mu\vert n::0,\phi\right)\delta_{j,n/2}}\nonumber \\
 & = & \sum_{n=1}^{\infty}2\mathcal{N}_{\alpha}^{2}\left|c_{n}\right|^{2}\left[\sum_{\mu=-n/2}^{n/2}\frac{\left[\partial_{\phi}p\left(\frac{n}{2},\mu\vert n::0,\phi\right)\right]^{2}}{p\left(\frac{n}{2},\mu\vert n::0,\phi\right)}\right]\nonumber \\
 & = & 2\mathcal{N}_{\alpha}^{2}\sum_{n=1}^{\infty}\left|c_{n}\right|^{2}n^{2},\label{eq:EC_CFI_DPC}
\end{eqnarray}
where we have introduced the simplified symbol $\partial/\partial\phi\equiv\partial_{\phi}$
and in the last equality we have used the result given by Eq.~\eqref{eq:CFI_NOON}.
It is remarkable that the $F_{{\rm EC}}$ is exactly equivalent to
the $F_{{\rm EC}}$ and also $\phi$-independent. Obviously, the results
given by Eqs.~\eqref{eq:QFI_EC} and \eqref{eq:EC_CFI_DPC} are equal,
which indicates that the DPC detection is also globally optimal for
EC states, meanwhile it has the same role as in the case of NOON states.

In order to compare with the result obtained in Ref.~\cite{Joo2011PRL}
but neglecting the effect of photon losses, we revisit the above phase
estimation problem by taking account of parity detection
as Joo \emph{et al.} done \cite{Joo2011PRL}.
We analytically derive the CFI with parity detection as follows (see
Appendix for detailed derivation)
\begin{eqnarray}
F_{{\rm EC}} & = & \frac{\left|2\mathcal{N}_{\alpha}^{2}\sum_{n=1}^{\infty}\left|c_{n}\right|^{2}n\sin\left(n\phi\right)\right|^{2}}{1-\left(2\mathcal{N}_{\alpha}^{2}\left[2\left|c_{0}\right|^{2}+\sum_{n=1}^{\infty}\left|c_{n}\right|^{2}\cos\left(n\phi\right)\right]\right)^{2}}.\label{eq:EC_CFI_parity}
\end{eqnarray}
It is explicitly simplified to $F_{{\rm EC}}=H_{{\rm EC}}$ in the
asymptotic limit $\phi\rightarrow0$, in the sense that the parity
detection is responsible of saturating its quantum analog $H_{{\rm EC}}$
given by Eq.~\eqref{eq:QFI_EC} only for small $\phi$. To view this
more clearly we plot in Fig.~\ref{fig:CFI} the CFIs with both DPC
and parity measurements as a function of $\phi$ for EC states according
to Eqs.~\eqref{eq:EC_CFI_DPC} and \eqref{eq:EC_CFI_parity}. It
can been seen that the value of the CFI with parity detection is lower
than that with PDC detection over the whole phase interval except
at the zero-point location, where they have merged satisfactorily.
Such a superiority of the PDC over parity can be intuitively understood
as parity results may be obtained from photon-counting results by
throwing away some information, while the opposite is not possible.
Thus this leads to the fact that the phase information contained in
the quantum state cannot be fully extracted by the parity measurement
in contrast to the PDC measurement under certain circumstances. It
is shown in Fig.~\ref{fig:CFI} that the sensitivity given by Joo
\emph{et al.} \cite{Joo2011PRL}
does not accessible with both the PDC and parity measurements, in
contrast, ours are accessible. We identify the reason underlying this
difference in the following subsection.

\subsection{Further discussion}

In above, we have assumed the phase shifting only
acting on the single arm of the interferometer (see Fig. \ref{fig:set-up}),
the same as the phase configuration considered in Ref.~\cite{Joo2011PRL}.
But the usual way to schematize a two-mode phase estimation scheme
involves the phase shifts of $\phi/2$ and $-\phi/2$ in the two arms,
instead of the single-phase shift $\phi$ in one arm. Let us further
compare the influence of the two-phase configurations on the phase
sensitivity. As assumed above, the single-phase shifting operation
is modeled by $U_{1}\left(\phi\right)=\exp\left(-ia^{\dagger}a\phi\right)$,
where we affix a subscript '1' to avoid confusion with the two-phase
shifting operation expressed by $U_{2}\left(\phi\right)=\exp\left[-i\left(a^{\dagger}a-b^{\dagger}b\right)\phi/2\right].$
By resorting to the Schwinger representation, the above two operations
can be reformulated as
\begin{eqnarray}
U_{1}\left(\phi\right)=\exp\Big[-i\Big(\frac{\hat{N}}{2}+J_{z}\Big)\phi\Big], & \quad{\rm and}\quad & U_{2}\left(\phi\right)=\exp\left(-iJ_{z}\phi\right),
\end{eqnarray}
where $\hat{N}=a^{\dagger}a+b^{\dagger}b$ is the total photon number
operator commuting with $J_{\xi}$ $\left(\xi=x,y,z\right)$. According
to Eq.~\eqref{eq:QFI_var}, it is straightforward to obtain the QFIs
with respect to $U_{1}\left(\phi\right)$ and $U_{2}\left(\phi\right)$
as follows
\begin{eqnarray}
H_{1}=\big\langle\Delta^{2}\hat{N}\big\rangle_{\psi}+{\rm Cov}\big(\hat{N},J_{z}\big)+H_{2}, & \quad{\rm and}\quad & H_{2}=4\left\langle \Delta^{2}J_{z}\right\rangle _{\psi},
\end{eqnarray}
with the covariance given by ${\rm Cov}\big(\hat{H},J_{z}\big)=\langle\hat{N}J_{z}\rangle_{\psi}-\langle\hat{N}\rangle_{\psi}\langle J_{z}\rangle_{\psi}$.
For probe states of fixed photon number, such as NOON states, it is
easy to find $H_{1}=H_{2}$, in sense that the two-phase configurations
are metrological equivalent.

However, if a probe state has the variable photon
number, such as EC states, the two configurations exhibit a different
performance in phase estimation problem since $H_{1}$ is always larger
than $H_{2}$ up to a non-vanishing part $\langle\Delta^{2}\hat{N}\rangle+{\rm Cov}(\hat{N},J_{z})$.
This is the reason why there were some works which claimed that a
higher phase sensitivity could be attained by encoding the phase through
$U_{1}$ \cite{Ono2010PRA,Joo2011PRL}. However, in order to realize
this advantage, it is necessary to establish a common reference beam
in experiments \cite{Molmer1997PRA,Bartlett2007RMP,Hyllus2010PRL,Jarzyna2012PRA},
otherwise it cannot be realized if without introducing additional
resources, such as the detection schemes mentioned above. This is
because a rare photon-counting detection cannot access the fraction
of sensitivity contributed by the coherence between states of different
photon number, namely $\langle\Delta^{2}\hat{N}\rangle+{\rm Cov}(\hat{N},J_{z})$.
This addresses the question raised at the end of
Sect.3(B).

Below, we demonstrate that the two configurations
are also metrological equivalent if the reference beam is absent.
Under such assumption, the general states of variable photon number
can be represented by an incoherent statistical ensemble of pure states
$\left|\psi_{N}\right\rangle $ with different number of photons as
follows \cite{Bartlett2007RMP,Hyllus2010PRL,Genoni2011PRL,Jarzyna2012PRA,Zhong2017PRA}
\begin{eqnarray}
\rho & = & \bigoplus_{N=0}^{\infty}p_{N}\left|\psi_{N}\right\rangle \left\langle \psi_{N}\right|.\label{eq:separable_state}
\end{eqnarray}
It can be obtained from a generic two-mode pure state $\vert\psi\rangle=\sum_{n,n^{\prime}}C_{n,n^{\prime}}\vert n,n^{\prime}\rangle$
by taking the phase-averaged operation \cite{Genoni2011PRL,Jarzyna2012PRA,Zhong2017PRA}.
Correspondingly, the state of Eq.~\eqref{eq:separable_state} is
identified with
\begin{eqnarray}
p_{N}=\sum_{n=0}^{N}\left|C_{n,N-n}\right|^{2}, & \quad{\rm and}\quad & \left|\psi_{N}\right\rangle =\frac{1}{\sqrt{p_{N}}}\sum_{n=0}^{N}C_{n,N-n}\left|n,N-n\right\rangle .
\end{eqnarray}
Notably, the resultant states are identical under $U_{1}\left(\phi\right)$
and $U_{2}\left(\phi\right)$, since the state of fixed photon number
remains unchanged up to a global phase by undergoing the evolution
with the Hamiltonian $\hat{N}/2$, such that
\begin{equation}
\rho\left(\phi\right)=\bigoplus_{N=0}^{\infty}p_{N}\big\vert\psi_{N}^{\left(1\right)}\left(\phi\right)\big\rangle\big\langle\psi_{N}^{\left(1\right)}\left(\phi\right)\big\vert=\bigoplus_{N=0}^{\infty}p_{N}\big\vert\psi_{N}^{\left(2\right)}\left(\phi\right)\big\rangle\big\langle\psi_{N}^{\left(2\right)}\left(\phi\right)\vert,
\end{equation}
with $\big\vert\psi_{N}^{\left(i\right)}\left(\phi\right)\big\rangle\equiv U_{i}\left(\phi\right)\left|\psi_{N}\right\rangle \left(i=1,2\right)$.
We thus explicitly derive the QFI of $\rho\left(\phi\right)$ as
\begin{eqnarray}
H_{\rho} & = & \bigoplus_{N=0}^{\infty}p_{N}4\left\langle \Delta^{2}J_{z}\right\rangle _{\psi_{N}}=H_{2},\label{eq:QFI_rho}
\end{eqnarray}
where the first equality is a result of the summability of the QFI
\cite{Helstrom1976Book,Fujiwara2001PRA} and the second equality
results from the identical expressions of $\vert\psi\rangle$ as $\vert\psi\rangle=\sum_{n,n^{\prime}}C_{n,n^{\prime}}\vert n,n^{\prime}\rangle=\bigoplus_{N=0}^{\infty}\sqrt{p_{N}}\left|\psi_{N}\right\rangle $.
The equality of Eq.~\eqref{eq:QFI_rho}
indicates that the two-phase configurations are identical for the
case of absence of a suitable reference beam, and the phase sensitivity
is invariant under $U_{2}\left(\phi\right)$ whether the reference
beam exists or not.
\begin{flushleft}
\begin{figure}[t]
\centering{}\includegraphics[scale=0.95]{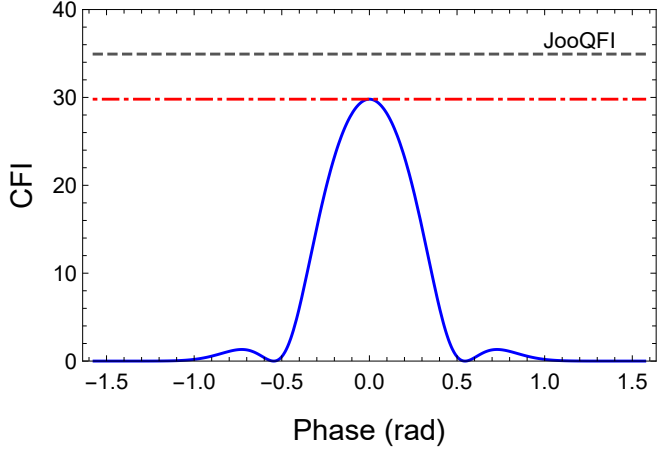}\caption{(Color online) The CFIs with DPC and parity measurements versus $\phi$
for the EC state with $\alpha=\sqrt{5}$. The blue curve corresponds
to the parity detection and the red dot-dashed line to the DPC measurement.
The black dashed line is given by $H_{\rm{EC}}^{J}=4\big[\mathcal{N}_{\alpha}^{2}\left(\vert\alpha\vert^{4}+\vert\alpha\vert^{2}\right)-\left(\mathcal{N}_{\alpha}^{2}\vert\alpha\vert^{2}\right)^{2}\big]$
derived by Joo \emph{et al.} in Ref.~\cite{Joo2011PRL}. \label{fig:CFI}}
\end{figure}
\par\end{flushleft}

\section{Conclusion}

In this paper, we have proposed even- and odd-orthogonality properties
of the Wigner D-matrix and provided an explicit
derivation of the two properties simply using the normalization condition
of quantum states. Resorting to these two properties, we identify
that the DPC detection is globally optimal for both NOON and EC states
in linear phase estimation. We find that for EC states this detection
exhibits outstanding performance than the parity detection when the
value of phase shift departing from the sweet point $\phi\sim0$.

Although only NOON and EC states are here taken as
examples, our newly derived properties can be also suitable for path-symmetric
pure states satisfying $\langle J_{z}\rangle_{\psi}=0$ \cite{Hofmann2009PRA,Lang2013PRL,Zhong2017PRA}.
This is an experimentally friendly condition for two-mode interferometric
phase estimation since most of probe states employed in experiments
belong to this family of states, such as the states created by injecting
coherent and squeezed vacuum states \cite{Caves1981PRD}, or two-mode
squeezed vacuum states \cite{Anisimov2010PRL} in a balanced beam
splitter. Moreover, we hope that there will be other applications
of our presented properties in addition to quantum metrology.

\section*{Acknowledgments}

This work was supported by the NSFC through Grant No. 12005106, the
Natural Science Foundation of the Jiangsu Higher Education Institutions
of China under Grant No. 20KJB140001 and a project funded by the Priority
Academic Program Development of Jiangsu Higher Education Institutions.
Y.B.S. acknowledges support from the NSFC through Grant No. 11974189.
L.Z. acknowledges support from the NSFC through Grant No. 12175106.

\section*{Data Availability Statement}
Data sharing is not applicable to this article as no
datasets were generated or analyzed during the current study.

\section*{Appendix: Phase estimation with parity detection \label{sec:AppendixA}}

  \makeatletter \renewcommand{\theequation}{A\arabic{equation}} \makeatother \setcounter{equation}{0}

In this appendix, we revisit the phase estimation problems discussed
in Sec. III in the main text by taking account of parity detection
\cite{Bollinger1996PRA,Gerry2000}. Assume a single-port parity detection
is carried out on the output mode $b$, which can be formulated as
$\Pi_{b}=\left(-1\right)^{b^{\dagger}b}=\exp\left(i\pi b^{\dagger}b\right).$
It accounts for distinguishing the states with even and odd numbers
of photons in a given output port. Specifically, the parity is assigned
as the value of $+1$ when the photon number of a state is even, and
the value of $-1$ if odd.

It has been demonstrated that the CFI in terms of parity detection
can be reformulated as \cite{Seshadreesan2013PRA,Zhong2021PRA}

\begin{eqnarray}
F & = & \frac{\left[\partial_{\phi}\langle\Pi_{b}\rangle\right]^{2}}{1-\langle\Pi_{b}\rangle^{2}}.\label{eq:CFI_parity}
\end{eqnarray}
It is apparent that this expression simply depends on the expectation
value of parity operator with respect to the output state to be detected.
For NOON states, one can easily obtain
\begin{eqnarray}
\langle\Pi_{b}\rangle_{{\rm NOON}} & = & \langle N::0,\phi\vert R_{y}^{\dagger}\left(\frac{\pi}{2}\right)\Pi_{b}R_{y}\left(\frac{\pi}{2}\right)\vert N::0,\phi\rangle=\begin{cases}
2, & N=0,\\
\cos\left(N\phi\right), & N\neq0.
\end{cases}\label{eq:parity_NOON}
\end{eqnarray}
By submitting Eq.~\eqref{eq:parity_NOON} into Eq.~\eqref{eq:CFI_parity},
it is straightforward to obtain $F_{{\rm NOON}}=N^{2}$, in the sense
that parity is optimal for NOON states as demonstrated in Sec.~III(A).
For EC states, the expectation value of $\Pi_{b}$ can be expressed
as the weighted linear combination of the expectations of $\Pi_{b}$
for NOON states with different photon numbers as
\begin{eqnarray}
\langle\Pi_{b}\rangle_{{\rm EC}} & = & 2\mathcal{N}_{\alpha}^{2}\sum_{n=0}^{\infty}\left|c_{n}\right|^{2}\langle\Pi_{b}\rangle_{{\rm noon}}=2\mathcal{N}_{\alpha}^{2}\left[2\left|c_{0}\right|^{2}+\sum_{n=1}^{\infty}\left|c_{n}\right|^{2}\cos\left(n\phi\right)\right],
\end{eqnarray}
where the last equality has used Eq.~\eqref{eq:parity_NOON}. Similarly,
submitting the above expression into Eq.~\eqref{eq:CFI_parity} yields
\begin{eqnarray}
F_{{\rm EC}} & = & \frac{\left|2\mathcal{N}_{\alpha}^{2}\sum_{n=1}^{\infty}\left|c_{n}\right|^{2}n\sin\left(n\phi\right)\right|^{2}}{1-\left(2\mathcal{N}_{\alpha}^{2}\left[2\left|c_{0}\right|^{2}+\sum_{n=1}^{\infty}\left|c_{n}\right|^{2}\cos\left(n\phi\right)\right]\right)^{2}}.
\end{eqnarray}

\end{document}